# Real-time volumetric MRI-guided MLC tracking: A feasibility study

Tianyu Yang, William P Donahue, Aniket Pramanik, Victor Murray, Can Wu, Ricardo Otazo, Grace Tang

*Memorial Sloan-Kettering Cancer Center, New York, NY, USA*

## Abstract

### Background and Purpose

The MR SIGnature MAtching (MRSIGMA) was recently proposed as a real-time free-breathing 3D MR imaging technique with an imaging latency of 148 ms. It is a powerful tool for intrafractional motion monitoring for disease sites such as those in the abdominal regions. This study presents a proof-of-concept MLC tracking system for dynamic adaptation of the radiation treatment based on the real-time target localization facilitated by MRSIGMA on a 1.5 T MR-Linac.

### Methods

An MLC tracking interface was developed for integration with MRSIGMA. The centroid position of a moving target was computed in real-time by MRSIGMA, which was then processed by a predictive filter to account for imaging and MLC adjustment latencies. The MLC leaves were repositioned by rigid translation. To validate the MLC tracking process, a 10 cm circular field was delivered to a moving phantom for a series of sinusoidal motions and patient motion. A clinical pancreas SBRT plan was also delivered with the patient motion trace.

### Results

For the circular field, geometric errors of <1.8 mm and overall system latency of < 125 ms were achieved for all motion traces. Compared to no tracking, MLC tracking remarkably improved the dosimetric accuracy for the circular field and the clinical pancreas plan, with an average Gamma passing rate increased from 87.1% to 99.9% based on a 3%/2mm criteria.

### Interpretation

A 3D MRI-guided MLC tracking technique was developed and validated on a clinical MR-Linac. Preliminary results demonstrated great dosimetric improvements with MLC tracking for treatments with moving target.

# Introduction

The current practice of motion management in radiation therapy utilizes treatment target expansion, patient breathing control, and radiation beam control. The choice of motion management method is generally dependent on the disease site and the patient's tolerability. For treatment regimen that deliver high doses in a few fractionations, such as stereotactic body radiation therapy (SBRT), motion management becomes even more critical, especially for abdominal diseases where respiration and gastrointestinal motions can cause anatomical changes to both the target and organs-at-risk (OARs)[1-3]. Abdominal compression and breath-hold techniques can minimize the target motion due to respiration, but reproducibility and residual motion within the breath-hold poses limitations[4,5]. In addition, patient experience with abdominal compression and breath-hold treatment can be uncomfortable and exhaustive. Respiratory gating is a more palatable motion management method, where patient is treated under free breathing. However, as radiation is only delivered when the target is within certain positional or phase range, treatment efficiency diminishes. In contrast, real-time tracking delivers radiation continuously as the beam is actively aligned to the target using dynamic multileaf collimator (MLC) motion, and treatment can be efficiently delivered with no additional cost to duty cycle and breath control[6,7].

Real-time MLC tracking has been implemented into the clinic and has demonstrated clinical benefits with potential enhancement in treatment outcome[8,9]. The first clinical implementation of MLC tracking was facilitated by an electromagnetic transponders system that provided real-time target position on conventional C-arm linac[10-12]. The electromagnetic transponders system provides internal target motion as the transponders are implanted into the patient. It has been reported to be an effective and efficient tool for real-time motion monitoring and has been widely used in the clinic[13-17]. However, the electromagnetic transponders introduce significant artifacts in MR images, posing issues in post-treatment/follow-up routine using MRI[18]. Other real-time motion monitoring methods such as fiducial-marker-based techniques using kV and/or MV imaging[19-23] have been explored and adopted in the clinic, but the concern of additional imaging dose and invasive procedure of implanting fiducial markers remains to be an ongoing debate.

With the advancement of MR-Linac technology, the superior soft tissue contrast in MR images offers a powerful tool for intra-fractional motion monitoring, of not just the target, but also OARs, during radiation treatment. In particular for treatment targets that are surrounded by soft tissues, such as those in the upper abdominal region, MR-guided radiation therapy facilitates online adaptive workflow that enables dose escalation while minimizing normal tissue toxicities[24,25]. For motion management, it has been reported that abdominal compression and gating show promising results using MR-cine imaging workflows[25-27]. Meanwhile, MLC tracking is yet to be commercially available, but several studies have supported its clinical feasibility[28-33]. These studies were based on 2D cine MRI to continuously monitor the intra-fractional motion of the target.

While interleaving cine MRI captures most of the target motion during clinical treatment[34-36], real-time 3D MRI is necessary to monitor the full deformable motions of the target and organs of interest. However, clinical implementation of real-time 3D MRI can be challenging due to the long acquisition and reconstruction time, robustness of motion binning, and presence of motion artifacts. Recently, a real-time 3D MRI technique, known as Magnetic Resonance SIGnature Matching (MRSIGMA), has been developed to fill this gap[37-40]. MRSIGMA provides high resolution volumetric images in the order of 1-2 mm in-plane and 2-4 mm through-plane with an imaging latency lower than 200 ms, enabling real-time intra-fractional motion monitoring during treatment on the MR-Linac. In this study, an MLC tracking software was developed and integrated with MRSIGMA for MRI-guided MLC tracking on a clinical 1.5 T MR-Linac.

# Methods

All experiments were performed on a clinical 1.5 T MR-Linac under a vendor-provided research environment (Unity, Elekta AB, Stockholm, Sweden).

## Real-time motion monitoring with MRSIGMA

The MRSIGMA workflow consisted of two scans: one for motion learning (non-real-time), and one for signature matching (real-time). During the motion learning scan, k-space data was collected and sorted into motion states by exploiting the self-navigating property of radial k-space data. Thirty motion states were generated with equal-amplitude binning of the estimated motion signal, and a deep learning framework, Movienet, was used for 4D MRI reconstruction[41]. The signature for each motion state was computed as the average z-projection for all the spokes within the motion state. The centroid of the target was extracted from each motion state to compile a dictionary of motion coordinates.

During the signature matching scan, each newly-acquired radial spoke generated a z-projection, which was matched to one of the motion signatures from the motion learning scan based on maximum correlation. The corresponding target centroid was selected as the output. With partial k-space sampling along $k_z$, the signature matching scan updated the 3D coordinates of the targets every 148 ms, enabling motion-resolved volumetric MR imaging. To reduce noise, an exponential moving average (EMA) filter was applied to the coordinates of the target.

## MLC tracking software

The MLC tracking software was built with two major components: a predictive filter and a command interface to the MLC controller on the linac. To compensate for system latencies from MR imaging, linac control communications, and mechanical movement of the MLC, the linear Kalman filter was used[42]. The Kalman filter attained the target centroid coordinates from MRSIGMA as inputs to predict target location at a later time. MRSIGMA provided a set of target coordinates $\{x_0, x_1, x_2, \ldots\}$ at times $\{t_0, t_0 + dt, t_0 + 2dt, \ldots\}$, where $dt = 148$ ms was the fixed time interval for MRSIGMA to acquire one spoke. After a learning period for parameter stabilization, the Kalman filter made a prediction of target coordinates upon receiving every incoming MRSIGMA input. The prediction length for the Kalman filter was $2 * dt$ to predict the MLC motion until the next available image was processed. Target position, velocity, and acceleration were used for the state vector. No observation noise was used since an EMA filter was already applied prior to the Kalman filter. The Kalman filter coordinates were converted to MLC leaf motions and the new MLC positions were sent to the MLC controller via the Elekta Digital Linac Interface (EDLI). The EDLI communication with the linac was up to 25 Hz, whereas the frequency was only 5 Hz for the clinically available iCom-Vx. The MLC tracking program checked the target displacement between the current and previous prediction and leaf velocities over $dt = 148$ ms were calculated. A control message was then sent to the MLC for each leaf to move at the calculated velocities and leaf velocities were updated upon each Kalman filter prediction. The update took effect at the next control cycle, which was every 40 ms. A script was developed to retrieve the actual MLC leaf positions from the linac in real-time at 25 Hz. The centroid of the recorded MLC aperture was calculated on the fly to monitor tracking performance. Figure 1 shows the schematics of the workflow between MRSIGMA, MLC tracking software, and the MR-linac.

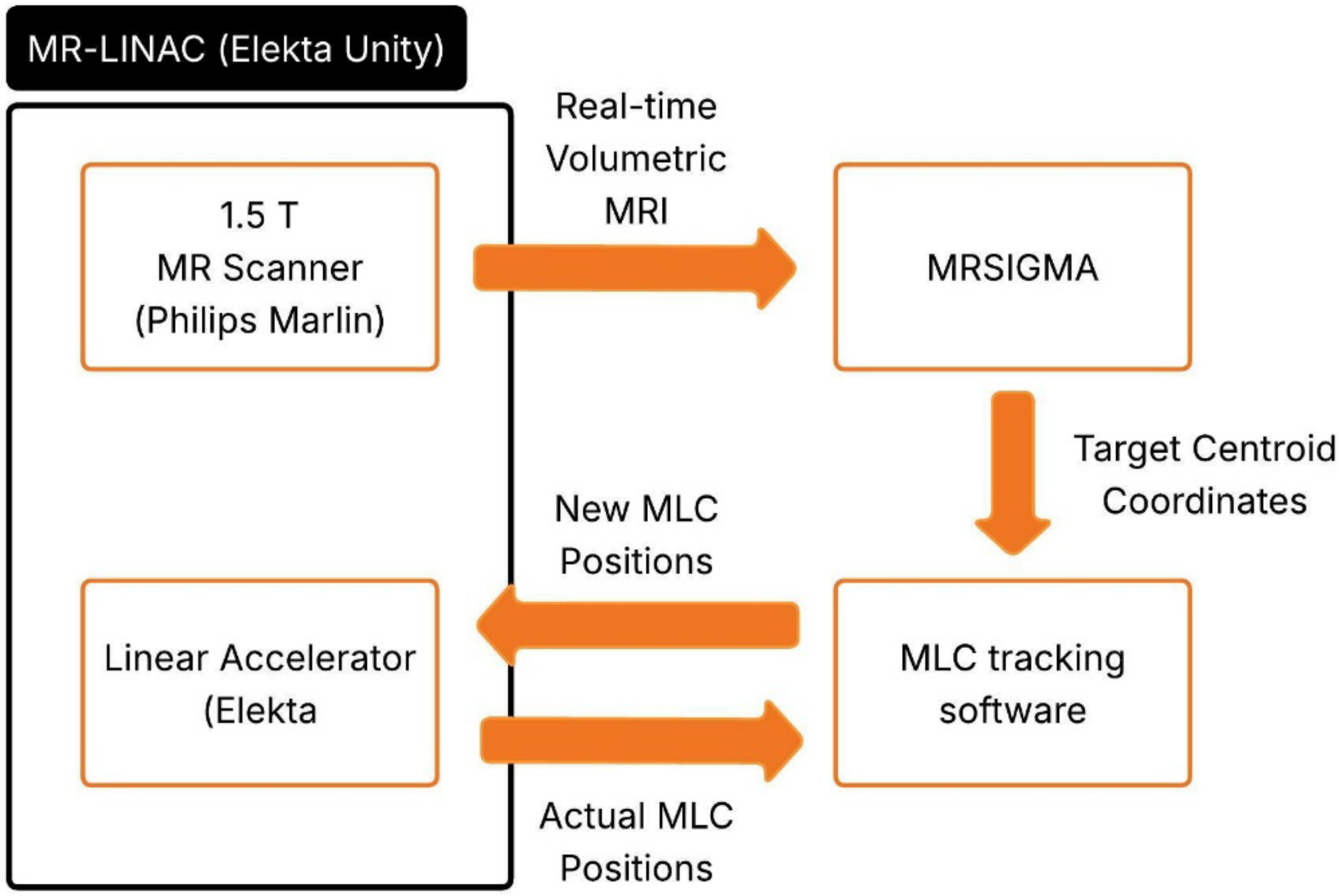

Figure 1. Overall schematics of the MRSIGMA-MLC tracking workflow.

## Validation testing

The MRSIGMA-guided MLC tracking workflow was evaluated using a custom phantom on an MR-compatible motion platform (QUASAR Motion MR Platform, Modus Medical Devices, London, Canada). The custom phantom was built in-house with a set of solid water slabs, with one of the slabs drilled and filled with copper sulfate to create a cylindrical MR-visible motion tracking surrogate with 2.5 cm in diameter and 5.5 cm in length. On the same solid water slab where the MR-visible target surrogate was located, two large cavities (radius 2.5 cm and length 27 cm) were drilled alongside the target and were filled with distilled water to increase the sensitivity for the MR scans. To minimize geometric distortions due to non-linearity in gradient fields, the target (hence the center of the custom phantom) was placed 2.5 cm below isocenter. The movement of the motion platform was along the superior-inferior direction, parallel to the MLC leaf travel.

MLC tracking performance was evaluated for geometric accuracy, system latency, and dosimetric accuracy. Geometric accuracy was assessed using a set of sinusoidal motion traces, with peak-to-peak amplitude of 15 mm and a range of periods of 5 to 8 s, and a motion trace derived from the liver dome for a pancreatic cancer patient using MRSIGMA (see Figure 2). For each motion trace, a static beam with a 10 cm circular aperture defined by the MLC was delivered at gantry 0 deg. The geometric accuracy was derived from comparing the recorded motion trace from the motion platform motor encoder (sample frequency: 1000 Hz), taken as ground truth, to the motion trace extracted from the retrieved actual/delivered MLC leaf positions during tracking (sample frequency: 25 Hz). For leaf positions recorded at each point in time, the centroid of the MLC aperture was computed and compared against that from the reference planned field. An MLC motion trace was subsequently composed with the displacements of the MLC aperture centroid. Synchronization of the MLC motion trace and the ground truth was performed implicitly, by recording the leaf positions on the same computer as the motion controller and by virtue of a <1 ms communication latency. Geometric accuracy was quantified by the root-mean-squared-error (RMSE) between the motion platform and MLC traces. The RMSE was defined as:

$$RMSE = \sqrt{\frac{1}{n}\sum_{i=1}^{n}(x_{MLC}(t_i) - x_{MP}(t_i))^2}$$

where $x_{MLC}(t_i)$ and $x_{MP}(t_i)$ were the MLC centroid and motion platform positions at time $t_i$, respectively, for an MLC motion trace recorded at times $\{t_1, t_2, \ \dots \ , \ t_n\}$ during beam-on. A beam-on signal from the MR-linac was recorded by the motion platform software to enable the selection of time points. For system latency, which included the time delay from target localization by MRSIMGA, processing of MLC tracking software for new MLC positions, and the actual adjustment of the MLC on the linac, was quantified by using cross-correlation between the motion platform trace and the MLC motion trace:

$$\hat{\tau} = \arg\max_{\tau \in \mathcal{T}} \sum_{i=1}^{n} \tilde{x}_{\mathrm{MLC}}(t_i)\, \tilde{x}_{\mathrm{MP}}(t_i + \tau)$$

where $\tilde{x}_{\mathrm{MLC}}(t_i) = x_{\mathrm{MLC}}(t_i) - \bar{x}_{\mathrm{MLC}}$ and $\tilde{x}_{\mathrm{MP}}(t_i) = x_{\mathrm{MP}}(t_i) - \bar{x}_{\mathrm{MP}}$ were the motion platform and MLC traces centered to zero.

The dosimetric performance of MRSIGMA-guided MLC tracking was assessed with film measurement (Gafchromic EBT4, Ashland Advanced Materials, LLC, NJ). Film was placed at 5 cm depth in the custom-made phantom to capture dose distributions in the coronal plane. Water was poured under and on top of the film to minimize air bubbles that could introduce unwanted dosimetric uncertainty from electron return effect. Prick marks were made on the film with four small pins that were built into the phantom to facilitate alignment/registration for film-to-film comparison. Film calibration was done by exposing a set of open 10 × 10 $cm^2$ fields with different levels of known dose. The circular field was delivered for all motion traces in three different settings: 1) reference delivery with no target motion; 2) delivery with motion and no tracking; and 3) delivery with motion and MLC tracking. A clinical 6.6/10 Gy × 5 fractions SBRT-IMRT pancreas plan was also delivered with the patient motion trace. After a minimum of 24 hours of self-developing time, the films were scanned with a flatbed scanner (Epson Expression 13000XL, Epson Seiko Corporation, Japan) and the calibration curve was derived with the FilmQA Pro software (Ashland Advanced Materials, LLC, NJ). Analysis was performed using an in-house software, where the deliveries with target motion were compared to the reference delivery with no motion. Gamma criteria of 3%/2mm was used for analysis, where dose agreement was based on 3% of the average dose within the region of interest defined by the 25% isodose line.

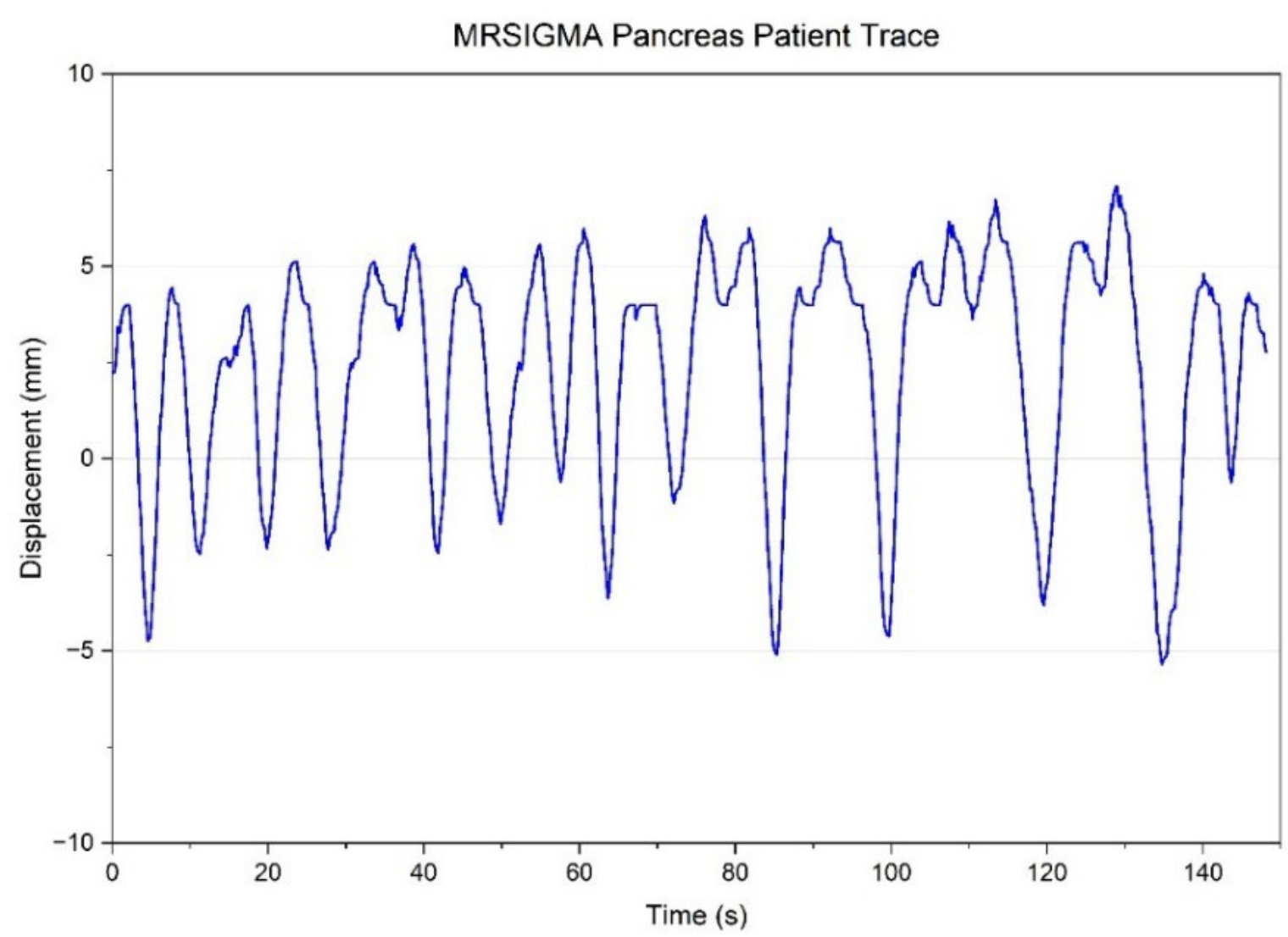

Figure 2. Patient motion trace acquired by MRSIGMA. The liver dome was used as the motion tracking target for this trace.

## Results

Table 1 summarizes the geometric accuracy and system latency results for MRSIGMA-guided MLC tracking based on a circular aperture for sinusoidal motion trace and a patient motion trace. For the sinusoidal motions, the geometric error increased with the speed of the moving target. This was also observed in the measured latency. For the patient trace, similar geometric accuracy and latency was yielded as the sinusoidal trace with a period of t = 8 s.

Table 1. Geometric accuracy and latency of MLC tracking for a circular field for different motion traces. The peak-to-peak amplitude for all sinusoidal motions was 15 mm.

| Motion | RMSE (mm) | Latency (ms) |
|---|---|---|
| Sine, T = 8 s | 0.79 | <40 |
| Sine, T = 6 s | 1.30 | 71 |
| Sine, T = 5 s | 1.79 | 125 |
| Patient trace | 0.70 | 70 |

For dosimetric accuracy, MLC tracking has greatly improved the agreement between the reference (i.e., expected) dose and the dose delivered to a moving target, with Gamma passing rates >96.5% achieved for all plans (see Table 1). In comparison, Gamma passing rates were below clinically acceptable level for deliveries to a moving target without MLC tracking[43]. As shown in Figure 3a, the Gamma passing rate was 84.5% with no tracking and local dose errors of >10% were observed in the clinical pancreas plan. With MLC tracking, the large dosimetric uncertainties were mitigated, as seen in Figure 3b.

| Motion | Plan | Gamma (3%/2mm) | |
|---|---|---|---|
| | | Motion no tracking | Motion with tracking |
| Sine, T = 8 s | Circular field | 87.9% | 99.9% |
| Sine, T = 6 s | Circular field | 87.6% | 96.5% |
| Sine, T = 5 s | Circular field | 87.2% | 100% |
| Patient trace | Circular field | 88.5% | 99.4% |
| Patient trace | Pancreas SBRT | 84.5% | 99.3% |

Table 2. Dosimetric accuracy of MLC tracking for a circular field and clinical pancreas SBRT plan.

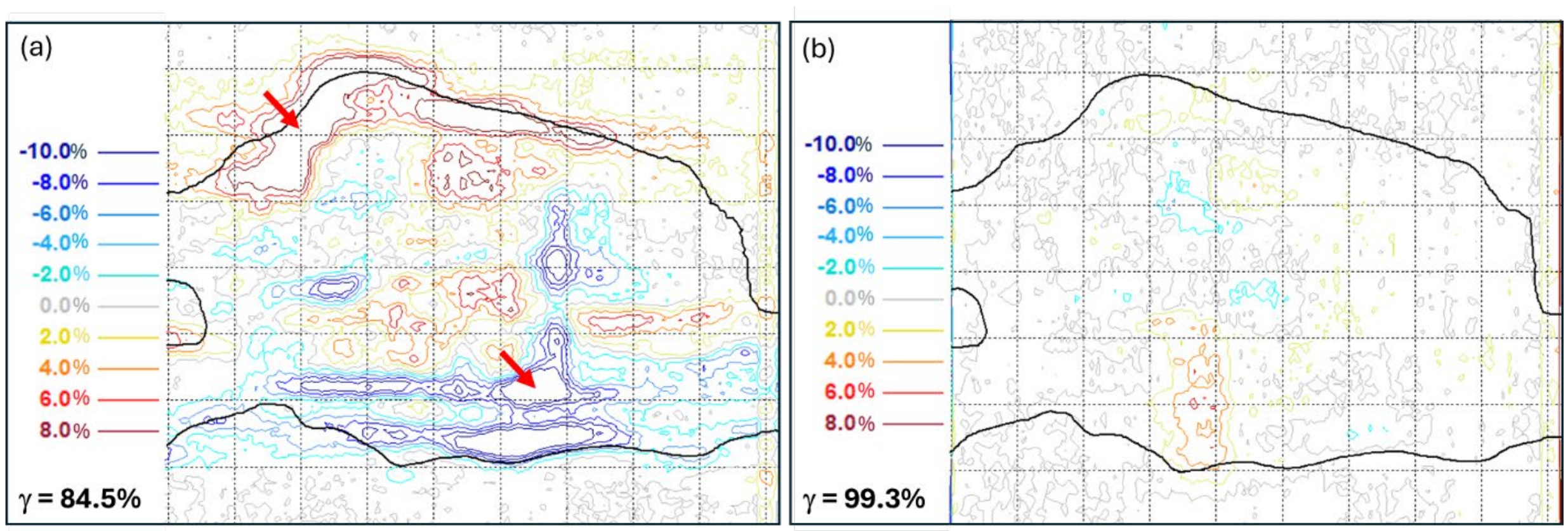

Figure 3. Dose difference distributions from film measurement between (a) motion with no MLC tracking – reference, and (b) motion with MLC tracking – reference, for the clinical pancreas plan. Local errors >10% were observed for delivery with no MLC tracking in multiple regions, as indicated by the red arrows.

## Discussion

This study demonstrates the first MLC tracking delivery with MRSIGMA, a real-time volumetric MR imaging technique. With most of the current motion management methods on the MR-Linac relying on the interleaved 2D cine MRI, this work presents an important step towards real-time adaptive therapy with active motion compensation under real-time 3D MR imaging guidance. For radiation treatment with a moving target, MLC tracking improved the dosimetric accuracy as demonstrated with the circular field and clinical plan measurement, rectifying the large dosimetric uncertainties exhibited in deliveries without MLC tracking. For all MLC tracking deliveries, geometric accuracy of 0.70 – 1.79 mm was achieved, with an overall system latency of <40 – 127 ms.

Although the system latency satisfies the current practice recommendations for MLC tracking[44], i.e., <500 ms, ongoing investigation is underway to minimize the time delay. On the other hand, there was some high frequency noise in the MLC motion traces that were reconstructed using the recorded MLC position data via EDLI, which partly contributed to the geometric errors. Compared to the ~1 mm geometric accuracy reported by Glitzner *et al* for a sinusoidal motion trace with higher peak-to-peak amplitude (i.e., 30 mm vs 15 mm) and shorter time period (3 s vs 5 s), the geometric accuracy attained in this current work warrants further improvement. This may include in-depth analysis of the accuracy and precision of the optical system that provides the actual MLC position information from the linac. Previous studies of MLC tracking on the MR-linac have not reported such issues, as the MLC motion reconstruction was performed using the MV images acquired by the electronic portal imaging device.

There are a couple of other areas for improvement for the MLC tracking software. First, analysis of MRSIGMA performance, localized the target with an accuracy of <1 mm for all the traces, but Kalman the filter over-estimated at the peaks and valleys of the target motion excursion. This demonstrates that the Kalman Filter requires further tuning to prevent over-estimation at points of rapid acceleration. These adjustments would include better tuning of internal parameters and the re-evaluation of model selection. Second, the current MLC positions and velocities need to be considered when computing their next step. This should produce more consistent delivery of the aperture adjustments.

Although the MLC tracking software was built for 2D motion tracking, for simplicity and ease of development, the preliminary work presented in this study was based on 1D motion and geometry-based tracking, i.e., motion compensation was performed using the centroid of the moving target. With the capability of providing 3D information of the target and OARs in real-time from MRSIGMA, future work includes developing anatomy-based tracking, which would mitigate over-irradiation of OARs that cannot be accounted for using geometry-based MLC tracking alone[45].

# Conclusions

A proof-of-concept real-time 3D MR-guided MLC tracking program was developed. Preliminary tests based on a static circular field and a clinical pancreas SBRT plan yielded high dosimetric accuracies. Ongoing works include refining the MLC tracking process, with further optimized parameters for the predictive filter and better dynamic control on MLC position adjustment using the real-time feedback from the linac.

# Acknowledgements

# Disclosure Statement

Elekta AB provided linear accelerator emulators and a customized Unity Operating System to enable this research.

# Data Availability Statement

Data availability is restricted due to ongoing vendor collaboration requirements.

# Author Contribution